# A Short-Term Voltage Stability Index and case studies

Wenlu Zhao

*Abstract*—The short-term voltage stability (SVS) problem in large-scale receiving-end power systems is serious due to the increasing load demand, the increasing use of electronically controlled loads and so on. Some serious blackouts are considered to be related to short-term voltage instability. In China, the East China Grid (ECG) is especially vulnerable to short-term voltage instability due the its increasing dependence on power injection from external grids through HVDC links. However, the SVS criteria used in practice are all qualitative and the SVS indices proposed in previous researches are mostly based on the qualitative SVS criteria. So a Short-Term Voltage Stability Index (*SVSI*), which is continuous, quantitative and multi-dimensional, is proposed in this paper. The *SVSI* consists of three components, which reflects the transient voltage restoration, the transient voltage oscillation and the steady-state recovery ability of the voltage signal respectively after the contingency has been cleared. The theoretical backgrounds and affected factors of these three components of *SVSI* are analyzed, together with some feasible applications. The verification of the validity of *SVSI* are tested through more 10,000 cases based on ECG. Additionally, a simple case of selecting candidate locations to install dynamic var using *SVSI* is presented to show its feasibility to solve the optimization problem for dynamic var allocation.

## I. INTRODUCTION

ACCORDING to the definition of short-term voltage stability (SVS) [1], it involves dynamics of fast acting load components such as induction motors, electronically controlled loads, and HVDC converters. Modern power systems are operating under more stressed situations due to the increasing load demands and electricity exchange between regions. Additionally, their dynamic responses are more complicated with the increasing use of induction motor loads (such as air-conditioners), electronically controlled loads, HVDCs and renewable power generations. As a result, the power systems are more vulnerable to short-term voltage instability. It is considered that part of the reason for the North American blackout in August 2003 may be related to SVS [2].

In China, most energy resources (such as coal, hydro and natural gas) are distributed in the west, while most loads are distributed in the east. So long-distance transmission is needed to deliver electric power from the west of China to the east of China. For instance, East China Grid (ECG) is a typical receiving-end system and receives power through HVDC links. According to the requirements for Chinese environmental protection, the increase of fossil-fuel generation is strictly controlled unless Combined Heat and Power (CHP) units. With the dependence on power injection from external grid increases year by year, the SVS in Yangtze River Delta is getting more serious. Therefore, the SVS in ECG, especially in Yangtze River-Delta, should be taken seriously.

Indices for describing voltage stability are required to evaluate the SVS. Currently the indices for describing transient stability can be divided into three categories: indices based on Transient Energy Function [3-6], Critical Clearing Time (CCT) [7] and indices based on voltage signals. The front two categories are most used in the study of rotor angle stability, while the latter one category is most used in the study of SVS. Some SVS criteria which used in practice are listed below [8-13]:

Western Electricity Coordinating Council (WECC):
- Voltage dip should not to exceed 25% at load buses or 30% at non-load buses and not to exceed 20% for more than 20 cycles at load buses under N-1 contingencies.
- Voltage dip should not to exceed 30% at any bus and not to exceed 20% for more than 40 cycles at load buses under N-2 contingencies.

Tennessee Valley Authority (TVA):
- Transmission system voltage should recover to 90% of nominal system voltage within 0.5 seconds of fault clearing under N-1 contingencies.
- Transmission system voltage should recover to 90% of nominal system voltage within 0.5 seconds of fault clearing or unit tripping under N-2 contingencies.

State Grid Corporation of China (SGCC):
Voltage should restore to 0.8p.u. within 10s after the fault has been cleared:

China Southern Power Grid (CSG)
Voltage should not drop to under a certain level such as 0.75p.u. lasting for longer than 1 s

Therefore, the SVS criteria used in practice can only determine the system is stability/instability, but they cannot reflect the degree of stability/instability. Furthermore, these criteria are merely dependent on the operating experiences and are lack of theoretical backgrounds. Therefore, the circumstance in which the criteria of WECC and TVA are in the same form but have different threshold occurs.

Due to the increasing complexity of modern power systems, the current SVS criteria can hardly satisfy the demand for the evaluation and optimization of SVS. Some indices to evaluate the SVS have been introduced in the previous researches [14-19], but they are mostly dependent on the above-mentioned SVS criteria and their theoretical backgrounds have not been discussed.

Therefore, a Short-Term Voltage Stability Index (*SVSI*), which is continuous, quantitative and multi-dimensional, is proposed in this paper to evaluate the SVS, and it is tested in the SVS evaluation of ECG to verify its effectiveness. The *SVSI* consists of three components, and each of them has theoretical backgrounds and affected factors. Moreover, the proposed *SVSI* can be additionally used in the researches on grid-partition, evaluation and optimization of dynamic vars and SVS evaluation using data-mining.

## II. SHORT-TERM VOLTAGE STABILITY INDEX

The Short-Term Voltage Stability Index (*SVSI*) proposed in this paper consists of three components: $SVSI_r$, $SVSI_o$, $SVSI_s$. These three components of *SVSI* reflects transient voltage restoration, transient voltage oscillation and ability to reach steady-state respectively.

The bus voltage signals considered in this paper are all normalized. Namely, if $v_0(t)$ is the voltage signal in transient process corresponding to a bus and $v_0(0)$ is the voltage of a bus before contingency, then the normalized voltage signal is:

$$v(t) = v_0(t)/v_0(0) \tag{1}$$

After normalization, all of the voltage signals starts at 1. The advantages of normalization will be stated below.

Based on the SVS criteria listed in Section I, a new definition of *transient voltage restoration* is proposed as follows: if the average of $v(t)$ is above 0.8pu in the time period which is from 9s to 10s after the contingency has been cleared, then the bus is considered as *transient voltage recovered*; otherwise is considered as *transient voltage unrecovered*.

The definition and algorithm of *SVSI* is different between the *transient voltage recovered* scenario and *transient voltage unrecovered* scenario. So $SVSI_r$, $SVSI_o$, $SVSI_s$ will be presented separately under these two scenarios.

### A. $SVSI_r$ under transient voltage recovered scenario

Two important arguments: $V_S$ and $T_{SVSIr}$ are needed to be defined before the calculation of $SVSI_r$.

The argument $V_S$ is defined to represent steady-state voltage after the transient process. Since the time for simulation is limited, the true value of $V_S$ is hard to obtain. Therefore, some logical judgments for approximation of $V_S$ is proposed.

The most common circumstance of the voltage signals after the contingency has been cleared is: the voltage firstly surges to a relative high status, then go through an oscillation damping period and finally comes to a steady-state. Therefore, the computational expression for $V_S$ under common circumstances is:

$$V_S = (\max(v(t))_{end} + \min(v(t))_{end})/2 \tag{2}$$

$\max(v(t))_{end}$ is the last peak of $v(t)$ and $\min(v(t))_{end}$ is the last valley of $v(t)$.

In addition, there are two special circumstances need to be considered while the fluctuation cycle and fluctuation amplitude of $v(t)$ cannot be recognized:

Firstly, if the slope of $v(t)$ tends to be gentle since the time when $v(t)$ reached the last extreme value, then:

$$V_S = v(T_{End}) \tag{3}$$

$T_{end}$ is the end time of simulation (i.e. end time of the SVS considered in this paper).

Secondly, if the slope of $v(t)$ tends to be steep since the time when $v(t)$ reached the last extreme value, then:

$$V_S = (\max(v(t))_{end} + \min(v(t))_{end})/2 \tag{4}$$

The definition of $T_{SVSIr}$ is: the moment when $v(t)$ firstly reaches $V_S$ during ramping periods after the contingency has been cleared. Under the most common circumstances, the computational expression for $T_{SVSIr}$ is:

$$\begin{aligned}&\min T_{SVSIr} \in (T_{Clear}, T_{End})\\ &s.t.\, v(t)\big|_{t=T_{SVSIr}} = V_S \\ &\exists \varepsilon > 0,\, \frac{dv(t)}{dt}\bigg|_{t \in (T_{SVSIr}-\varepsilon, T_{SVSIr}+\varepsilon)} \geq 0 \\ &\exists t_0 \in (T_{SVSIr}-\varepsilon, T_{SVSIr}+\varepsilon),\, \frac{dv(t)}{dt}\bigg|_{t=t_0} > 0 \end{aligned} \tag{5}$$

$T_{Clear}$ is the moment in simulation when contingency is cleared. In addition, there is a special circumstance need to be considered in which the inequality $v(t) > V_S$ holds from $T_{Clear}$ to $T_{End}$, thus:

$$T_{SVSIr} = T_{Clear} \tag{6}$$

Based on the arguments defined above and simulation result, the computational expression of $t_{SVSIr}$ is:

$$SVSI_r = \int_{T_{Flt}}^{T_{SVSIr}} |v(t) - V_S|\, dt \tag{7}$$

$T_{Flt}$ is the moment when contingency starts.

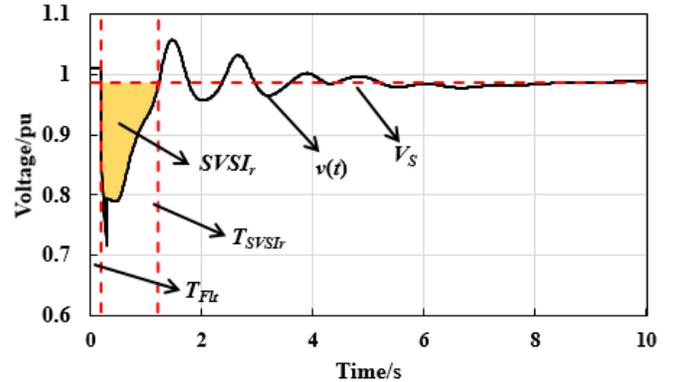

Fig 1 The schematic of $SVSI_r$ under transient voltage recovered scenario

In Fig 1, the area of yellow region is $SVSI_r$. $v(t)$ maintains nearly post-contingency steady-state ($V_S$) after $t_{SVSIr}$, but experiences a continuous low-voltage stage before $t_{SVSIr}$. In this paper, the time period from $t_{Flt}$ to $t_{SVSIr}$ is defined as *transient voltage restoration stage* and the time period from $t_{SVSIr}$ to $t_{End}$ is defined as *after transient voltage restoration stage*.

$SVSI_r$ is the transient voltage restoration component of *SVSI*. It reflects the dynamic reactive power balance and the transient voltage restoration in the short time after the contingency has been cleared. It is closely related to the dynamic response of induction motors (such as AC air-conditioners), over-excitation limits of generators, dynamic vars and HVDCs. The $SVSI_r$ reflects the key factors of SVS, and the greater the value is, the greater is the possibility of short-term voltage instability.

During the process of transient voltage restoration, the induction motors and HVDCs consumes a massive of reactive power and thus threaten the voltage stability. On the contrary, the over-excitation of generators and the dynamic vars such as STATCOMs and SVCs can rapidly send out reactive power to support the SVS. After the transient voltage restoration process, the variation of operating states of the induction motors and the HVDCs are not threatening because their consumption of reactive power returns to the normal level and thus the threaten to SVS is reduced.

The induction motors consume a massive of reactive power when they are operating under low voltage or stalled status. The induction motors threaten the SVS during *transient voltage restoration stage* because the bus remains under-voltage. The operating states of the induction motors depend on the balance between the electromagnetic torque ($Te$) and the mechanical torque ($Tm$). $Te$ is proportional to the square of bus voltage and $Tm$ usually increases with the speed of the motor. The bus voltage reaches the post-contingency steady-state level at the end of *transient voltage restoration stage* under *transient voltage recovered* scenario, so $Te$ recovers to the post-contingency steady-state level as well. So $Te$ is sufficient to maintain the balance with $Tm$ after *transient voltage restoration stage* and thus the variation of operating states of the induction motors are not threatening. Coupled with the effect of damping, the induction motors will return to steady-state eventually.

The HVDCs consume a massive of reactive power in the restoration process after commutation failures, which are harmful to the SVS. The operating states of the HVDCs are closely related to the voltage at the coupling point with AC system ($V_{AC}$) and the extinction angles (γ). When γ decreases, which may cause by a voltage dip, to a threshold value, the commutation failure of the HVDC will occurs. The bus remains under-voltage during *transient voltage restoration stage*, so the commutation failure of the HVDC may occur and thus threaten the SVS. Regardless of whether the commutation failure occurred in *transient voltage restoration stage*, the HVDC will operate normally after *transient voltage restoration stage* under *transient voltage recovered* scenario. Because the HVDC operates normally at post-contingency steady-state and its control strategy of commutation failure prevention after it has been restored from commutation failure allows it to have a relative strong ability against voltage disturbance.

To reduce *SVSIr* and enhance SVS, we can focus on: modification of induction motors, adjustment of HVDC controllers, optimize AVR parameters and increase dynamic var reservation.

*B. SVSIo under transient voltage recovered scenario*

The definition of $V_{SVSIo}(t)$ is needed before the calculation of $SVSI_o$. $V_{SVSIo}(t)$ reflects the mid-line of oscillation in $v(t)$ as shown in Fig 2. The oscillation of $v(t)$ can be considered in the area of power system small signal analysis, as the transient process can be studied in the area of small signal stability analysis when the voltage recovers to a certain level and the system tends to a steady-state after the contingency has been cleared. So, the aperiodic or low frequency fluctuations (usually lower than 0.1 Hz [20]) component of $v(t)$ is retained in $V_{SVSIo}(t)$, while the high frequency fluctuations (usually lower than 2 Hz [20]) of $v(t)$ are considered as noises as shown in Fig 5. The computational expression for $SVSI_o$ is:

$$SVSI_o = \int_{t_{Clear}}^{t_{End}} |v_{TVSIo}(t) - v(t)| dt \quad (8)$$

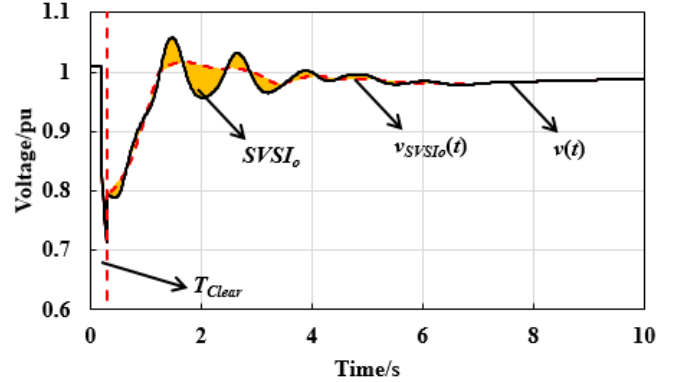

Fig 2 The schematic of $SVSI_o$ under transient voltage recovered scenario

In Fig 2, the area of yellow region is $SVSI_o$. It is the transient voltage oscillation component of $SVSI$, which reflects the balance of active power between generators and loads and the damping ability of the system. It is closely related to dynamic responses of high-gain fast-excitation devices [21], generator speed control systems [22] and PSSs. The greater the $SVSI_o$ is, the slower the oscillation attenuates after the contingency has been cleared, and thus the greater is the possibility of short-term voltage instability.

To reduce $SVSI_o$ and enhance SVS, we can focus on optimizing of the dynamic control of generators and performance of PSSs.

*C. SVSIs. under transient voltage recovered scenario*

The definition of $T_{Stable}$ is need to be defined before the calculation of $SVSI_s$. $T_{Stable}$ is the moment when $v(t)$ reaches a post-contingency steady-state. Since the simulation time is limited, the true value of $T_{Stable}$ is hard to obtain. Therefore, some logical judgments for approximation of $T_{Stable}$ is proposed.

The computational expression for $T_{Stable}$ under the most common circumstances is:

$$\min T_{Stable} \\ s.t. \forall t \in [T_{Stable}, T_{End}], |V_S - v(t)| < V_{wTh} \quad (9)$$

$V_{wTh}$ is the amplitude threshold of voltage fluctuation/wave for determining steady-state. Refers to the requirements for voltage steady-state judgement in [21], $V_{wTh}$ is set to be $0.01 p.u.$

In addition, there are two special circumstances to be considered. Firstly, if a continuous oscillation with increasing amplitude is observed in $v(t)$, then:

$$T_{Stable} = T_{End} \quad (10)$$

Secondly, if $\forall t \in [T_{Clear}, T_{End}], |V_S - v(t)| < V_{wTh}$ holds, then:

$$T_{Stable} = T_{Clear} \quad (11)$$

Based on the arguments defined above and the voltage signal, the computational expression for $SVSI_s$ is:

$$SVSI_s = (v(0) - V_S) \times (t_{Stable} - t_{Clear}) \quad (12)$$

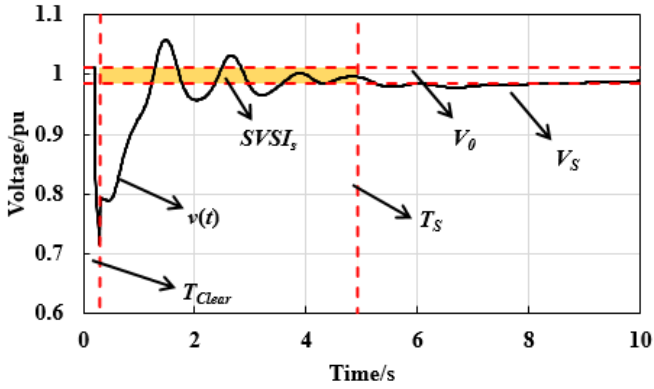

Fig 3 The schematic of $SVSI_s$ under transient voltage recovered scenario

In Fig 3, the area of yellow region is $SVSI_s$. It is the steady-state recovery ability component of $SVSI$, which reflects the speed to reach post-contingency steady-state and voltage level of bus at post-contingency steady-state. It is closely related to the topology and operating mode (such as: tripping transmission lines, load-shedding, switching shunt compensations) of the system. The greater the $SVSI_s$ is, the slower the post-contingency steady-state reaches. and the lower the voltage of bus is at post-contingency steady-state, and thus the greater is the possibility of short-term voltage instability.

To reduce $SVSI_s$ and enhance SVS, we can focus on the modification of post-contingency operating mode of the system.

### D. SVSI under transient voltage unrecovered scenarios

The main difference of $SVSI_r$, $SVSI_o$ and $SVSI_s$ under *transient voltage unrecovered scenario* is the different definitions of $V_S$, $T_{SVSIr}$, $T_{Stable}$ and $SVSI_r$. The definitions of rest of the arguments and indices are the same as the *transient voltage recovered scenario*.

The new definitions of $V_S$, $T_{SVSIr}$, $T_{Stable}$ and $SVSI_r$ are:

$$V_S = v_{SVSIo}(T_{End}) \quad (13)$$
$$T_{SVSIr} = T_{End} \quad (14)$$
$$T_{Stable} = T_{End} \quad (15)$$
$$SVSI_r = \int_{T_{Flt}}^{T_{TVSIr}} |v(t) - v(0)| dt \quad (16)$$

The new schematics for $SVSI_r$, $SVSI_o$ and $SVSI_s$ are shown in Fig 4-6.

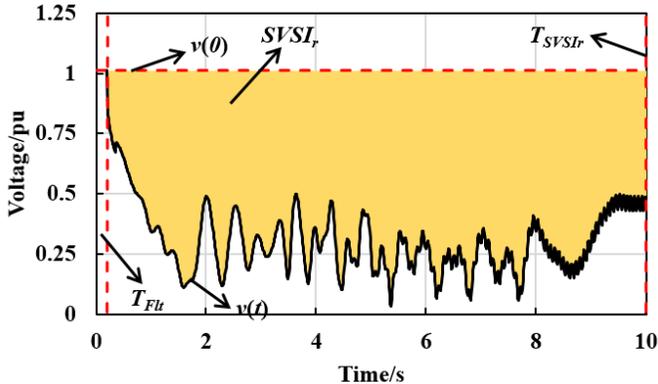

Fig 4 The schematic of $SVSI_r$ under transient voltage unrecovered scenarios

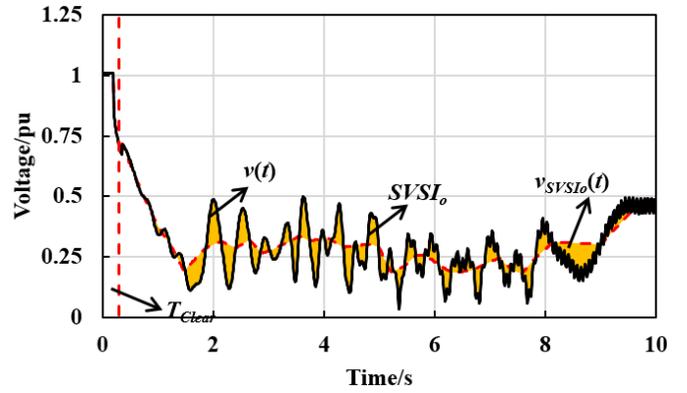

Fig 5 The schematic of $SVSI_o$ under transient voltage unrecovered scenarios

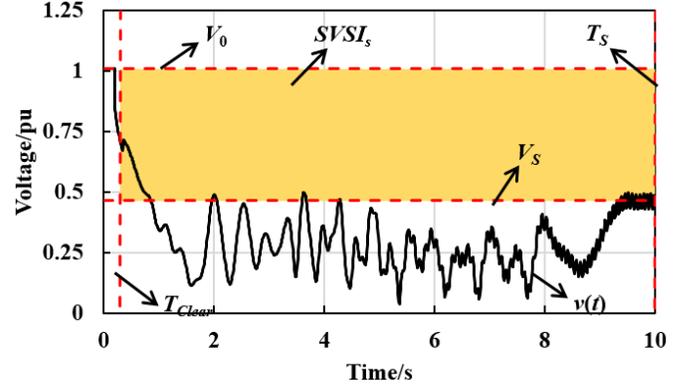

Fig 6 The schematic of $SVSI_s$ under transient voltage unrecovered scenarios

### E. Some Discussions

*1) Motivation of Normalization*

The major factor of most short-term voltage instability cases is the continuous under-voltage at buses. So when suffering from a same voltage dip, the SVS of the buses whose voltage is originally in a lower level is worse, while by contrast the buses whose voltage is originally in a higher level is better. Normalization increases the $SVSI$ corresponding to the buses whose voltage is originally lower than base level, and decreases the $SVSI$ corresponding to the buses whose voltage is originally higher than base level, which makes the result of SVS evaluation more reasonable.

*2) Probable application of SVSI*

   a) Grid-Partition According to $SVSI$

Utilizing $SVSI$ to describe the response of voltage after the contingency has been cleared, thus the grid-partition can be made according to the $SVSI$ corresponding to buses under the set of credible contingencies. The grid-partition results are mainly affected by the topology and the set of credible contingencies of the system, which usually change little for a long time. Therefore, the grid-partition result can be used in SVS evaluation until the next change of topology or set of credible contingencies of the system, thus reducing the computational amount.

   b) Evaluation and Optimization of Dynamic VAR Reserve

$SVSI$ can quantitatively evaluate the SVS, so the sensitivity function between dynamic var and $SVSI$ can be used for the evaluation and optimization of dynamic var reserve.

   c) Real-Time Evaluation of SVS based on Data-Mining

The theoretical backgrounds and the affected factors of the *SVSI* are stated above. Utilizing these affected factors as input and the *SVSI* as output, the supervised learning artificial neural network (ANN) can be constructed. The constructed ANN model can evaluate the SVS much faster than Time Domain Simulation, so it can be used in on-line evaluation of SVS.

In summary, the *SVSI* consists of three components, which can reflect the key features of voltage signals after the contingency has been cleared. Moreover, each of the components has theoretical backgrounds and affected factors. Therefore, *SVSI* is a comprehensive criterion to evaluate the SVS and it can be applied to various researches.

## III. CASE STUDIES

The validity of *SVSI* are verified based on ECG system, which contains 3,435 buses and 5,838 transmission lines. The typical operating mode in summer is chosen for analysis, because the load level is summer is high and thus the SVS of the system is vulnerable. More than 10,000 cases have been carried out for the verification, but only four typical and one practical instance are shown due to limited space.

### A. Case I: Modify the proportion of induction motor loads

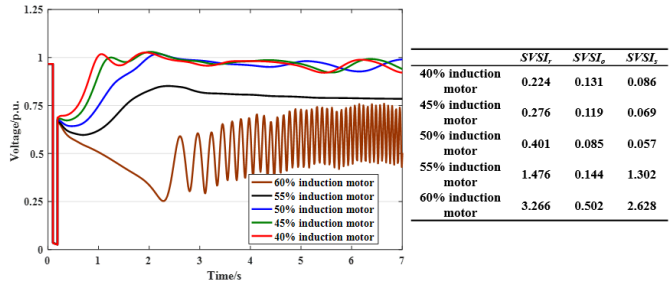

Fig 7 The voltage signals and *SVSI* under the scenarios with different proportion of induction motor loads

Fig 7 shows the voltage signals and the *SVSI* as the proportion of induction motors in loads changes. From the voltage signals on the left, the restoration ability of voltage signals becomes weaker as the proportion of induction motor increases. This conforms to the statement about the impact of induction motors on SVS as above-mentioned. From the *SVSI* on the right, $SVSI_r$ increases significantly as the proportion of induction motor increases. As a contrast, $SVSI_o$ and $SVSI_s$ change little when the proportion of induction motor is less than 50%.

From the voltage signals, when the proportion of induction motor increases to 55%, the voltage signal is nearly unstable; when the proportion increase to 60%, the short-term voltage instability occurs. From the *SVSI*, the $SVSI_r$, $SVSI_o$ and $SVSI_s$ are all increased significantly when the proportion of induction motor increased to more than 55%. So if all of the three components of *SVSI* increases significantly, the voltage signal may nearly unstable or already unstable. This conclusion remains correctness in the following cases.

### B. Case II: Modify the installations of dynamic var

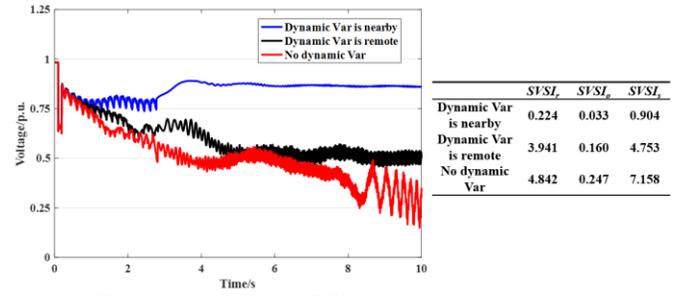

Fig 8 The voltage signals and *SVSI* under the scenarios with different configurations of dynamic var

Fig 8 show the voltage signals and the *SVSI* as the installation of dynamic var changes. From the voltage signals on the left, the restoration ability of voltage signals becomes better when the dynamic var is installed. Moreover, the voltage restoration ability of voltage signals is better when the dynamic var locates near the measurement bus in contrast to the dynamic var locates remote from the measurement bus. From the *SVSI* on the right, all of the three components increases as the dynamic var arrangement changes in the sequence of no dynamic var, dynamic var is remote and dynamic var is nearby.

### C. Case III: Modify the configuration of PSS

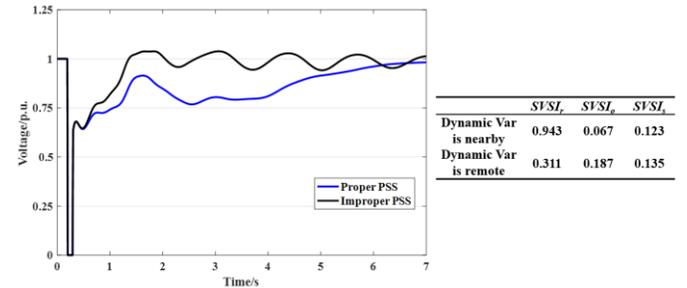

Fig 9 The voltage signals and *SVSI* under the scenarios with different configuration of PSS

Fig 9 shows the voltage signals and the *SVSI* as the configuration of PSS changes. From the voltage signals on the left, the oscillation of voltage signals is more obvious when the system is configured with proper PSS in contrast to when the system is configured with improper PSS. From the *SVSI* on the right, the $SVSI_o$ is smaller when the PSS is proper. The reason for the $SVSI_r$ is larger under the 'Proper PSS' case may be that the PSS restricts the response of fast-excitation to prevent the sustained oscillation after voltage restoration.

### D. Case IV: Modify the strategy of under-voltage load-shedding

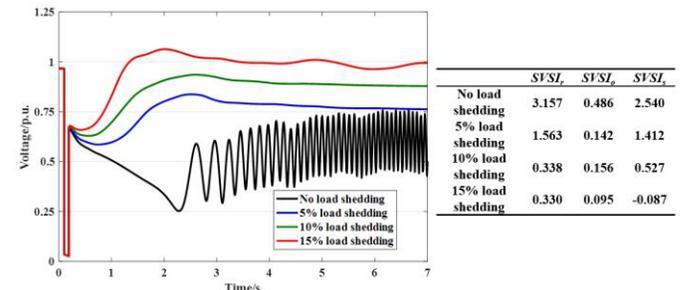

Fig 10 The voltage signals and *SVSI* under the scenarios with different strategies of under-voltage load-shedding

Fig 10 shows the voltage signals and *SVSI* as the strategy of under-voltage load-shedding changes. From the voltage signals on the left, the ability of voltage signals to reach steady-state and the voltage level at post-contingency steady-state is better as the proportion of load-shedding increases. From the *SVSI* on the right, the *SVSI*$_s$ decreases as the proportion of load-shedding increases.

*E. Case V: select installation locations for dynamic var*

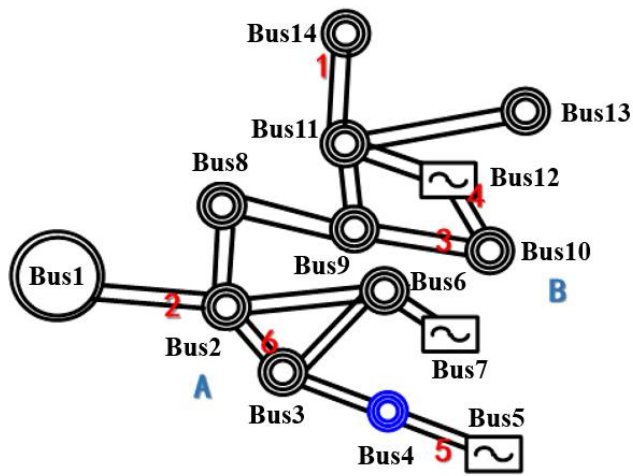

Fig 11  Diagram of part of ECG

Fig 11 is the diagram of a part of ECG. The buses are anonymous due to confidential considerations. This part locates at the end of East China Gird, which connects to the rest part only through Bus1. The red numbers represents the candidate locations of contingencies and the blue letters represents the candidate location to install dynamic vars.

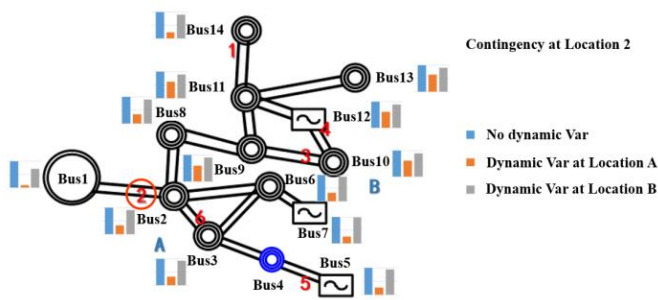

Fig 12  Effect of dynamic var when contingency occurred at Location 2

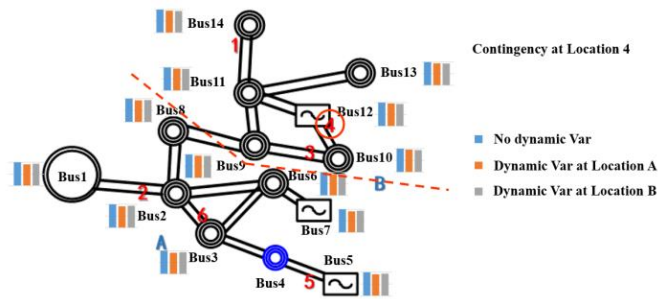

Fig 13  Effect of dynamic var when contingency occurred at Location 4

Fig 12 and Fig 13 show the composite value of *SVSI* when contingency occurred at two of the candidate locations. Under the case that the contingency was occurred at location 2, all of the *SVSI* in this part are much smaller when the dynamic var is installed at Location A (closer to contingency location) in contrast to when the dynamic var is installed at Location B. However, under the case that contingency occurred at location 4, only part of the *SVSI* in this part is smaller when the dynamic var is installed at Location B (closer to contingency location) in contrast to when the dynamic var is installed at Location A. Therefore, Location A is a better choice for dynamic var installation in contrast to Location B. Furthermore, the optimal locations to install dynamic vars can be solved based on the above-mentioned procedure.

## IV. CONCLUSION

A continuous, quantitative and multidimensional index (*SVSI*) is presented in this paper for evaluation of SVS. *SVSI*r, *SVSI*o and *SVSI*s are the three components of *SVSI*, which reflects the transient voltage restoration, the transient voltage oscillation and the steady-state recovery ability of the voltage signal respectively after the contingency has been cleared. The theoretical backgrounds and affected factors of the three components are analyzed in detail to reveal the wide applicability of *SVSI*.

More than 10,000 cases based on ECG have been carried out to verify the effectiveness of *SVSI* and five of them are presented in this paper. The presented cases shown that:

1) The *SVSI* can reflect the key characteristics of the voltage signal after the contingency has been cleared.

2) The three components of *SVSI* can be adjusted through the modifications of their affected factors respectively.

3) The *SVSI* can be applied to determining the optimal locations to install dynamic vars.

Moreover, the *SVSI* can also be applied to grid-partitioning, evaluation and optimization of dynamic var reserve, Real-Time evaluation of SVS based on Data-Mining and so on. These are also the following research contents of this study.